\documentstyle[aps,eqsecnum,amssymb,12pt]{revtex} 
\begin{document} 

\tighten
 
\title{ Can one increase the luminosity of a Schwarzschild black 
  hole?} 
 
\author{Avraham E. Mayo\thanks{Electronic mail: Mayo@cc.huji.ac.il}} 
 
\address{\it The Racah Institute of Physics, Hebrew University of 
  Jerusalem,\\ Givat Ram, Jerusalem 91904, Israel}

\date{\today} 
 
\maketitle 

\begin{abstract} 
We illustrate how Hawking's radiance from a Schwarzschild black hole
is {\it modified} by the electrostatic self-interaction of the emitted
charged particles. A W.K.B approximation shows that the probability
for a self-interacting charged particle to propagate from the interior
to the exterior of the horizon is increased relative to the
corresponding probability for neutral particles.  We also demonstrate
how the electric potential of a charged test object in the black
hole's vicinity gives rise to pair creation.  We analyze this
phenomenon semiclassically by considering the existence of the
appropriate {\it Klein region}.  Finally we discuss the possible
energy source for the process.
\end{abstract} 
 
\section{Introduction} 
 
The aim of this work is to study various effects that may take place 
when a charged object propagates on a background Schwarzschild 
spacetime.  As elucidated by Vilenkin\cite{Vilenkin} and corroborated 
by Smith and Will\cite{Smith} and Lohiya\cite{Lohiya}, in contrast to 
the situation in flat spacetime, in the presence of a Schwarzschild 
black hole the {\it self-field} of the object, makes a nontrivial 
contribution to the object's energy as measured at infinity. 
Alternatively, this phenomenon may be interpreted as black hole 
polarization; the test object polarize the black hole so that image 
charges are formed on the black hole horizon. It was shown by various 
authors \cite{BekMayo} that this effect is of unquestionable 
importance in black hole thermodynamics. One may think of two 
additional {\it distinct} effects that may take place on account of 
black hole polarization. The first is the emission of self-interacting 
charged particles. Here the black hole is {\it isolated}. The 
temperature of the black hole is assumed to be high enough so that the 
lightest massive charged particles may be emitted. The second effect 
is pair creation due to vacuum polarization. The black hole is allowed 
to interact with an \textit{exterior } charged test object, and the 
potential generated by the charge is taken into account. Both 
processes seems to have the effect of increasing the luminosity of the 
black hole in a sense that will be clarified below. 
 
In the following sections we investigate these two effects.  First we
show that once the temperature of the Schwarzschild black hole $\sim
1/M$, is equal or greater than the rest-mass $\mu$ of a
self-interacting charged particle ( $\mu\lesssim 1/M$), the emission
of these particles is more probable than the emission of similar
neutral particles by a factor $\exp(\pi e^2)$, where $e$ is the charge
(unless the contrary is stated we use ``natural units'' in which
$\hbar =c=G=k=1$ throughout).This is a direct result of the
self-interaction of the charged particles. In analogy with the
increased thermionic emission from a hot filament placed in a
repelling electrostatic field, which has the effect of reducing the
potential barrier at the edge of the metal, the effective
electrostatic self-field experienced by the charges has the effect of
increasing the emission \cite{Schiffer}. In order to demonstrate this
we compute in Sec.\ref{hawking} the probabilities for finding
particles in the exterior and interior of the horizon using the
appropriate wave equation coupled to an effective electromagnetic
self-field. It turns out that for a Schwarzschild black hole, the
chemical potential is lowered (for both signs of charge ) by
$e^2/8M$. As pointed out before by several researches
\cite{Vilenkin,Lohiya}, this suggests that the emission of charged
particles would increase relative to the emission of neutral
particles. We calculate an approximation to the absorption coefficient
(``greybody factor'') for the problem. We find that when the energy of
the emitted particles is close to $e^2/8M$, the absorption coefficient
is of the form $a(\omega,\mu,e^2/8M)(\omega-e^2/8M)$ plus small
corrections. This means that in the energy range $\mu<\omega< e^2/8M$,
the black hole should start to superradiate .
 
Thus far we have discussed the emission of charged self-interacting 
particles. In Sec.\ref{klein} we change the settings and discuss the 
consequences of placing a charged test object in the {\it exterior} of 
a Schwarzschild black-hole. We illustrate how the field generated by 
the charged particle gives rise to pair production.  We analyze the 
problem semiclassically and find the energy range for the existence of 
a Klein region. It was previously thought that this energy range 
does not exist in the case of an {\it isolated} nonrotating neutral 
black-hole \cite{Deruelle}. 
 
In Sec. IV we address the problem of possible energy sources for the
process. Thermodynamical arguments suggests that if the black-hole is
too massive, the entropy carried away by the pairs is not sufficient
to compensate for the entropy lost by the black-hole. Accordingly we
look into the dynamics of the test object, and investigate the role it
may play in the supply of energy for the process. Finally in Sec. V we
summaries our finidng. 
 
\section{Particle emission rate.} 
\label{hawking} 
 
We take now a semiclassical approach to the calculation of the
transmissions probabilities. We will use the complex path method which
was recently advanced by various authors
\cite{Pady,Damour:therm,Maulik}. Specifically Srinivasan and
Padmanabhan\cite{Pady} proposed a derivation of Hawking radiation
without using the Kruskal extension as in \cite{HartleHawk}. The
point, of course, is that the Schwarzschild coordinates possess a
coordinate singularity at the horizon. This bad behavior of the
coordinates manifests itself as singularity in the expression for the
semiclassical propagator near the horizon and a specific prescription
to bypass it must be provided. The action functional may be
constructed using the Hamilton-Jacobi method in the appropriate
coordinates. It will be shown how the analytic continuation used in
complex path method gives the result that the probability for
particles to be found in the exterior of the horizon is not the same
as the probability for particles to be found in the interior of the
horizon. The ratio between these probabilities is of the form, $P_{\rm
interior} = \chi \exp(-\beta \varepsilon)\ P_{\rm exterior}$, where
$\varepsilon$ is the energy of the particles and $\beta = 8\pi M$ is
the inverse Hawking temperature. $\chi$ is a factor of the form
$\chi=\exp(\pi e^2)$, where $e$ is the charge of the test
particle. This implies that the usual thermal distribution of neutral
particles is modified by the addition of what may be interpreted as a
charge dependent chemical potential.
 
\subsection{Transmission probabilities} 
 
In a Schwarzschild spacetime consider a minimally coupled test scalar
field $\Phi$ with mass $\mu$ coupled to the electromagnetic self-field
$A_\alpha$ generated by the charge $e$. The scalar field propagates in
the metric
\begin{equation} 
ds^2 = -(1-2M/r)dt^2 +(1-2M/r)^{-1} dr^2 +r^2\; d\Omega^2, 
\label{metric} 
\end{equation} 
where $d\Omega^2=d\theta^2+\sin\theta^2 d\varphi^2$, and satisfies 
\begin{equation} 
\left(\nabla_\alpha-{\imath\over\hbar} e A_\alpha\right)\left( 
  \nabla^\alpha-{\imath\over\hbar} e A^\alpha\right) \Phi 
={\mu^2\over\hbar^2} \Phi. 
\label{kg} 
\end{equation} 
$A_\alpha$, linear in $e$, is the self-potential whose source is the 
object itself\cite{Vilenkin,Smith,Lohiya}. 
\begin{equation} 
A_\alpha=-{e M\over 2 r^2} \delta^t_\alpha. 
\label{self} 
\end{equation} 
$A_\alpha$ is divergence free accurate to $O(e^2)$.  Using this and 
expanding the left hand side of Eq.~(\ref{kg}), we obtain, 
{  
\begin{equation} 
{1\over r^2}{ \partial _r}\left[r^2(1-2M/r){\partial _r \Phi}\right] 
-( 1-2M/r)^{-1}\left({\partial _t}-{\imath\over\hbar}e A_t\right)^2\Phi-{1\over 
  r^2}\hat {L}^2\Phi= {\mu^2 \over \hbar^2} \Phi , 
\label{kgr} 
\end{equation}
} 
where $\hat {L}^2$ is the usual squared angular momentum operator with 
eigenvalues $L^2=\hbar ^2 l(l+1)$. 
 
Since the problem is a spherically symmetric one, we put $\Phi 
=\exp((\imath/\hbar) {\cal S}(r,t)) Y^m_l(\theta, \varphi)$ where 
${\cal S}$ is a function which will be expanded in powers of $\hbar$. 
Substituting in the above equation, we obtain, 
{ 
\begin{eqnarray} 
 &&\left[ -(1-2M/r)^{-1} \left( {\partial _t {\cal S} } 
-e A_t\right)^2  +(1-2M/r)\left(\partial_r {\cal S} 
\right)^2 +\left(\mu^2+{L^2/ r^2}\right) \right] 
\nonumber \\ 
&&+{\hbar\over\imath}\left[-(1-2M/r)^{-1}\partial^2_t{\cal S}+{1\over 
    r^2}\partial_r(r^2(1-2M/r))\partial_r{\cal 
    S}+(1-2M/r)\partial^2_r{\cal S}\right]=0. 
\nonumber \\ 
\label{seqn1} 
 \end{eqnarray} 
}
 Expanding ${\cal S}(r,t) = {\cal S}_0(r,t) + \left({\hbar / 
 \imath}\right) {\cal S}_1(r,t) +\left({\hbar / \imath}\right)^2 {\cal 
 S}_2(r,t) +\cdots $, substituting into Eq.~(\ref{seqn1}) and 
 neglecting terms of order $\hbar/\imath$ and higher, we find to 
 lowest order, 
{ 
\begin{equation} 
 -{(1-2M/r)^{-1}} \left( {\partial_t {\cal S}_0} -e A_t\right)^2 
 +(1-2M/r)\left( {\partial_r {\cal S}_0} \right)^2 
 +\left(\mu^2+{L^2/ r^2}\right) =0. 
\label{seqn2} 
\end{equation} 
}
This is just the Hamilton-Jacobi equation satisfied by a charged 
particle of mass $\mu$ and charge $e$ moving in the spacetime 
(\ref{metric}) and interacting with the potential $A_\alpha$ as given 
in Eq.~(\ref{self}). The solution to the above equation is 
{ 
\begin{equation} 
{\cal S}_0(r,t) = -\varepsilon t \pm \int^r dr (1-2M/r)^{-1} 
\sqrt{(\varepsilon+e A_t)^2 - (1-2M/r)(\mu^2 +L^2/ 
r^2)}, \label{ssol1} 
\end{equation} 
}
where $\varepsilon$ is a constant identified with the energy. 
 
The semiclassical kernel ${\cal K}(r_2,t_2; r_1,t_1)$ for the particle 
to propagate from $(t_1,r_1)$ to $(t_2,r_2)$ in the saddle point 
approximation can be written down immediately: 
\begin{equation} 
{\cal K}(r_2,t_2; r_1,t_1) = {\cal N} \exp\left( {\imath\over \hbar} 
{\cal S}(r_2,t_2; r_1,t_1) \right). 
\end{equation} 
${\cal N}$ is a suitable normalization constant, and ${\cal S}$ is the 
action functional satisfying the classical Hamilton-Jacobi equation, 
namely 
\begin{equation} 
{\cal S}(r_2,t_2; r_1,t_1)={\cal S}_0(r_2,t_2)-{\cal S}_0(r_1,t_1). 
 \label{eqn:ssol3} 
\end{equation} 
The sign ambiguity in ${\cal S}_0 $ is related to the ``outgoing'' 
($\partial_r {\cal S}\,>0$) or ``ingoing'' ($\partial _r{\cal S} 
\,<0$) nature of the particle.  As long as points $1$ and $2$, between 
which the transition amplitude is calculated, are on the same side of 
the horizon (i.e.  either exterior or interior to the horizon), the 
integral in the action is well defined and real.  But if the points 
are located on opposite sides of the horizon, then the integral does 
not exist due to the divergence of the integrand at $r=2M$. Therefore, 
in order to obtain the probability amplitude for crossing the horizon 
we have to give a prescription for evaluating the integral. The 
prescription used by Srinivasan and Padmanabhan\cite{Pady}, and 
adopted here, is to take the contour defining the integral to be an 
infinitesimal semicircle {\it above} the pole at $r=2M$ for outgoing 
particles on the left of the horizon and ingoing particles on the 
right. Similarly, for ingoing particles on the left and outgoing 
particles on the right of the horizon (which corresponds to a time 
reversed situation of the previous cases), the contour should be an 
infinitesimal semicircle {\it below} the pole at $r=2M$. 
 
Consider, therefore, an outgoing particle ($\partial_r {\cal S}\,>0$) 
at $r=r_1<2M$. The squared modulus of the amplitude for this particle 
to cross the horizon gives the probability to find the particle in the 
exterior of the horizon. The contribution to ${\cal S}$ in the ranges 
$(r_1,2M-\delta)$ and $(2M+\delta,r_2)$ is real.  Therefore, choosing 
the contour to lie in the upper complex plane, 
\begin{eqnarray} 
{\cal S}_{{r_1}_{\overrightarrow{\;\; \,_{2M}\; \; }} r_2} &=& 
-\lim_{\delta \to 0} \int^{2M+\delta}_{2M-\delta} { \varepsilon+e 
A_t\over 1-2M/r}dr + \; ({\rm real \; part}) \nonumber \\ &=& 
\imath 2\pi M(\varepsilon+A_t(2M))+ ({\rm real \; part}) 
\end{eqnarray} 
where $A_t$ is evaluated at $r=2M$. The minus sign in front of the 
integral corresponds to the initial condition that $\partial 
_r{\cal 
  S} \,>0$ at $r=r_1<2M$.  The same result is obtained when an ingoing 
particle ($\partial_r {\cal S} \,<0$) is considered at $r=r_1<2M$. 
The contour for this case must be chosen to lie in the lower complex 
plane.  The amplitude for this particle to cross the horizon is the 
same as that of the outgoing particle due to the time reversal 
symmetry obeyed by the system. 
 
Consider next, an ingoing particle ($\partial_r {\cal S} \,<0$) at 
$r=r_2>2M$.  The squared modulus of the amplitude for this particle to 
cross the horizon gives the probability to find the particle in the 
interior of the horizon. Choosing the contour to lie in the upper 
complex plane, we get, 
\begin{eqnarray} 
{\cal S} _{{r_1}_{\overleftarrow{\; \; \,_{2M}\; \; }} r_2} &=& 
-\lim_{\delta \to 0} \int^{2M-\delta}_{2M+\delta} { 
\varepsilon+eA_t\over 1-2M/r} dr+ \; ({\rm real \; part}) 
\nonumber \\ &=& -\imath 2\pi M (\varepsilon+eA_t(2M))  + ({\rm 
real \; part}) 
\end{eqnarray} 
where as before $A_t$ is evaluated at $r=2M$. The same result is 
obtained when an outgoing particle ($\partial_r {\cal S}\,>0$) is 
considered at $r=r_2>2M$.  The contour for this case should be in 
the lower complex plane and the amplitude for this particle to 
cross the horizon is the same as that of the ingoing particle due 
to time reversal invariance. 
 
Taking the modulus square to obtain the probability, and 
substituting for $A_t$ from Eq.~(\ref{self}), we get, 
\begin{equation} 
\pmatrix{\textit{probability for a particle} \cr \textit{with
energy}\,\, \varepsilon \cr \textit{to be found at $r>2M$}}= \exp{(-8
\pi M \varepsilon+\pi e^2)} \pmatrix{\textit{probability for a
particle} \cr \textit{with energy}\,\, \varepsilon \cr \textit{to be
found at $r<2M$}}.
 \label{ssys} 
\end{equation} 
This suggest that it is more likely for a particular region to gain
particles than lose them.  Therefore we must interpret the above
result as saying that the probability of emission of particles is not
the same as the probability of absorption of particles. Furthermore,
this result implies that the flux of particle at infinity in this case
would be greater by a factor $\exp({\pi e^2})$ than the flux of
neutral particles of the same mass and spin. However, it should be
noted that the increase factor is in fact negligiblely small when one
considers the emission of electrons. In that case the numerical factor
is equal to $\exp({\pi {e^2/( \hbar c)}})\sim\exp({\pi /137})\approx
1.023$. This result obviously cast doubt on the astrophysical
importance of the electrostatic self-interaction.

Note that the method presented here for the calculation of the
modulation factor of Hawking's radiance is not specific to the
electrostatic self-potential. It would also apply for an {\it
external} electrostatic potential (as opposed to the {\it internal}
electrostatic self-potential). Provided that the external potential is
analytic in the neighborhood of the horizon, the thermal radiation is
modulated by varying the magnitude of the external potential. The
modulation factor is of the form $\exp(e A^{\em ext}_t(2M)/T_H)$,
where the external potential, $A^{\em ext}_t$ is evaluated at the
horizon and $T_H$ is the Hawking temperature.
 
Recently van Putten\cite{Putten} discussed the magnetic analog of this
phenomenon. He showed that a rotating black-hole produces
electron-positron outflow when immersed in a magnetic field. The
outflow is driven primarily by a coupling of the spin of the black
hole to the fermionic wave-function.  The source for the magnetic
field is assumed to be exterior. He computed, in the WKB
approximation, the superradiant amplification of scalar waves confined
to a thin equatorial wedge around a Kerr black-hole and found it to be
higher than for radiation incident over all angles. However, Aguirre
\cite{Aguirre} has recently presented calculations of both spin-0
(scalar) superradiance (integrating the radial equation rather than
using the WKB method) and spin-1 (electromagnetic/magnetosonic)
superradiance in van Putten's wedge geometry, and showed that in
contrast to the scalar case, spin-1 superradiance is weaker in the
wedge geometry. So that, as with the electrostatic self-interaction
case, the astrophysical significance of the effect is questionable.
 
\subsection{ The greybody factor} 
 
It follows from the previous subsection that Hawking emission of 
charged self-interacting particles from the black-hole gives rise to a 
flux at infinity 
\begin{equation} 
{d^2 n\over d\omega dt}={1\over 2\pi}{\Gamma \over e^{8\pi M 
    (\omega-e^2/8M)}-1}, 
\label{flux} 
\end{equation} 
where $\omega$ is the energy of the particle at infinity, $e$ is the 
charge of the particles, $8\pi M$ is the inverse of Hawking 
temperature, and $\Gamma$ is the relevant absorption factor, the so 
called ``greybody factor''.  From the equation above it is apparent 
that the term $-e^2/8M$ serves as a chemical potential in the problem; 
intriguingly, this chemical potential is the same for both particles 
and antiparticles (usually chemical potential has opposite sign for 
particles and antiparticles). It may be seen that this chemical 
potential enhances the emission of (positive or negative) charged 
particles over that of neutral particles with the same mass and spin. 
Positive and negative charged particles are treated equally. 
 
It turns out that two different situations occur for $\omega>e^2/8M$ 
and $\omega<e^2/8M$. To see this, expand the right hand side of 
Eq.~(\ref{flux}) around $\omega=e^2/8M$. Then, we get that the flux at 
infinity is 
\begin{equation} 
{d^2 n\over d\omega dt}\approx {1\over 2\pi}{\Gamma \over 8\pi M 
(\omega-e^2/8M)}. 
\end{equation} 
It is now evident that for $\omega>e^2/8M$ the flux is positive. 
However for $\omega<e^2/8M$ the flux seems to be negative, with a 
singularity at $\omega=e^2/8M$ ! None the less, one shouldn't be 
perplexed by this. It is just that we have failed to take into account 
the energy dependence of the absorption factor $\Gamma$. Obviously, we 
must show that the the point of transition occurs at a zero of 
$\Gamma$, namely the expansion of the absorption factor around 
$\omega=e^2/8M$ must begin with 
\begin{equation} 
\Gamma\approx a(\omega,\mu,e^2/8M)(\omega-e^2/8M)+\cdots, 
\label{aproxgamma} 
\end{equation} 
where $a(\omega,\mu,e^2/8M)$ is finite function of its variables. 
 
Therefore, we turn to calculate $\Gamma$ to first approximation. 
Making the ansatz $\Phi=e^{-\imath \omega t} Y^m_l(\theta, \varphi) 
f(r)/r$ and substituting into Eq.~(\ref{kgr}) we obtain an effective 
Schr\" odinger equation 
\begin{eqnarray} 
&&-\hbar^2{d^2 f\over {dr^*}^2}+V(r^*)f=0, \nonumber \\ && 
\matrix{V(r^*)&=\left(1-{2M\over r}\right )\left(\mu^2+{ L^2\over 
r^2}+{\hbar^2 2M\over r^3}\right)\cr &-\left(\omega-{e^2 M\over 2 
r^2}\right)^2.\hfill} \nonumber \\ 
\label{sch} 
\end{eqnarray} 
Here $r^*=r+2M\log(r-2M)$ is Wheeler's ``tortoise'' coordinate 
\cite{MTW}.  We are interested in asymptotic solutions of the equation 
in the {\it far region} $r^*\rightarrow\infty\;(r\rightarrow\infty)$ 
and in the {\it near region}, $r^*\rightarrow -\infty\;(r\rightarrow 
2M)$. 
 
Taking the limit of $V(r^*)$ in the {\it far region}, 
$r^*\rightarrow\infty\;(r\rightarrow\infty)$, we find that the 
equation has the form 
\begin{equation} 
-\hbar^2{d^2 f\over {dr^*}^2}+(\mu^2-\omega^2) f=0. 
\label{r->inf} 
\end{equation} 
For $\mu>\omega$ this equation has exponentially decaying and 
diverging solutions. Obviously, the diverging solution is unacceptable 
on physical grounds, while the decaying solution is associated with a 
particle trapped in the effective potential well. For $\mu<\omega$ 
Eq.~(\ref{r->inf}) has an ingoing and an outgoing wave solutions. 
Since we are actually dealing with a scattering problem, we 
concentrate on those solutions: 
\begin{equation} 
f(r\rightarrow\infty)=e^{-\imath\sqrt{\omega^2-\mu^2}r^*/\hbar}+{\cal 
A} e^{\imath\sqrt{\omega^2-\mu^2}r^*/\hbar}. 
\end{equation} 
Here ${\cal A}$ is a constant to be determined later. 
 
In the {\it near region}, $r^*\rightarrow\infty\;(r\rightarrow 2M)$, 
we find that Eq.~(\ref{sch}) has the limiting form 
\begin{equation} 
-\hbar^2{d^2 f\over {dr^*}^2}-(\omega-e^2/8M)^2f=0. 
\end{equation} 
The Matzner boundary condition \cite{Matzner} that the physical 
solution be an ingoing wave, as appropriate to the absorbing character 
of the horizon, selects the solution of the above equation to be 
\begin{equation} 
f(r\rightarrow 2M)={\cal B} e^{-\imath(\omega-e^2/8M)r^*/\hbar}, 
\end{equation} 
where ${\cal B}$ is a constant. 
 
${\cal A}$ and ${\cal B}$ may be determined by matching $f$ and $f'$ 
of the solutions in the {\it far and near regions} at some point in 
the intermediate region $2M\ll r\ll \infty$. Doing so we find 
\begin{eqnarray} 
&& {\cal A}={\omega-e^2/8M-\sqrt{\omega^2-\mu^2}\over 
\omega-e^2/8M+\sqrt{\omega^2-\mu^2}}e^{-2{\imath\over 
\hbar}\sqrt{\omega^2-\mu^2}r^*_m}, \nonumber \\ 
&& {\cal B} = {2\sqrt{\omega^2-\mu^2}\over 
\omega-e^2/8M+\sqrt{\omega^2-\mu^2}}e^{-{\imath\over\hbar}\left(\sqrt{\omega^2-\mu^2}- 
(\omega-{e^2/ 8M})\right)r^*_m}, 
\nonumber \\ 
\label{amp} 
\end{eqnarray} 
where $r^*_m$ is the matching point.  The absorption cross-section may 
now be obtained using the method of fluxes. The flux in our 
one-dimensional effective problem is 
\begin{equation} 
{\cal F}={\hbar\over 2\imath}\left(f^*\partial_r f-c.c.\right). 
\end{equation} 
The absorption probability is the ratio of the incoming flux at the 
horizon to the incoming flux at infinity, 
\begin{eqnarray} 
\Gamma&=&{{\cal F}_{\cal H}\over {\cal F}_\infty^{incoming}}= 
\nonumber \\ 
&=&{4 
\sqrt{\omega^2-\mu^2}\over (\omega-e^2/8M 
-\sqrt{\omega^2-\mu^2})^2}(\omega-e^2 /8M). 
\nonumber \\ 
\label{grey} 
\end{eqnarray} 
The matching point $r^*_m$ has disappeared from the expression for 
$\Gamma$. This means that at this order of approximation the matching 
point may be chosen arbitrarily as long as it is in the range $2M\ll\ 
r^*_m\ll\infty$. 
 
Note that for $e^2/8M\approx \omega$ this expression agree with our 
supposition (\ref{aproxgamma}). The drastic difference between the two 
regimes, $\omega<e^2 /8M$ and $e^2/8M<\omega$ shows up clearly if we 
consider the limit where the effective temperature of the black-hole 
tends to zero. If the self-energy of the particles is neglected, the 
rate of particles creation goes then to zero.  In our system the rate 
goes also to zero if $e^2/8M<\omega$. But in the range 
$\mu<\omega<e^2/8M$ the rate of particle creation tends to 
$-\Gamma/2\pi$. We are in the domain of black-hole superradiance of 
charged particles. 
 
\section{Pair production near a Schwarzschild black-hole} 
\label{klein} 
 
While Hawking radiance relates closely to dynamical spacetimes with 
horizons, pair creation does not. One may inquire for the 
circumstances in which a {\it neutral} black-hole is involved in the 
spontaneous production of pairs of oppositely charged particles. 
Technically, pair production can take place when the conditions for 
the existence of a ``generalized ergosphere'' (a region where orbits 
with negative total energy exist) are fulfilled. In this case, on the 
level of classical particles, a Penrose process \cite{Penrose} can 
take place. On the level of waves mechanics similar conditions must be 
fulfilled in order for ``superradiant'' scattering of waves obeying 
the Klein-Gordon equation to occur. The corresponding phenomenon that 
occurs then is the so called {\it Klein paradox}\cite{Sakurai}. 
However, all this is known {\it not} to happen when the black-hole is a 
neutral static and isolated one. But, it turns out that once the black 
hole is allowed to interact with an {\it exterior} charged test 
object, this may give rise to pair production.  Technically, this 
means that the conditions for the occurrence of a  Klein paradox 
are obeyed. Then it is possible to calculate in a W.K.B. approximation 
the transmission coefficient through the {\it Klein region} separating 
the positive from the negative states of the corresponding 
Klein-Gordon equation. This gives the probability for pair creation. 
 
Using the Hamilton-Jacobi equation Eq.~(\ref{seqn2}), one may derive 
the equation of motion of a classical particle of mass $\mu$ and 
charge $e$ on the background metric (\ref{metric}): 
\begin{eqnarray} 
&& \left({dr\over d\tau}\right)^2=(\varepsilon-{\cal 
E}_+(r))(\varepsilon-{\cal E}_-(r)) , \nonumber \\ && {\cal 
E}_\pm(r)={e^2 M\over 2r^2}\pm\sqrt{\left(1-{2M\over 
r}\right)\left(\mu^2+{L^2\over r^2}\right)}. 
 \label{epm} 
\end{eqnarray} 
Here $r$ and $\tau$ are the radial coordinate and proper time of the
particle respectively, and ${\cal E}_+$ and ${\cal E}_-$ are the
effective potentials for the positive and negative energy solutions,
respectively. The classical bound states in the potentials ${\cal
E}_\pm$ are the classical limit of the ``resonances'' of a quantum
field satisfying the Klein-Gordon equation (\ref{kg}) written in the given
background metric. Positive energy states, $\varepsilon_+>{\cal E}_+$,
correspond to a positive probability density, and therefore describe
particles of energy $\varepsilon$.  Negative energy states,
$\varepsilon_-<{\cal E}_-$ correspond to a negative probability
density and therefore describe antiparticles of energy $-\varepsilon$.
When there is a crossing of the $\varepsilon_+$ and the
$\varepsilon_-$ states, the probability density has a variable sign.
We are then in a Klein paradox region.
 
The transmission coefficient $T^2$ through the potential barrier 
separating the positive from the negative energy states is 
proportional to the probability for an incident particle to create a 
pair of particles. The transmission coefficient can be computed using 
the W.K.B approximation to Eq.~(\ref{sch}): 
\begin{eqnarray} 
  T^2&=&\exp\left(-\zeta\right), \nonumber \\ 
  \zeta&=&2\int^{\alpha_1}_{\alpha_2}{dr\over 1-2M/r}\sqrt{V(r)}. 
\end{eqnarray} 
Here $\zeta$ may be identified as the opacity of the barrier against 
pair creation. $\alpha_1$ and $\alpha_2$ are the zeroes of the 
effective potential $V(r)$ defined in Eq.~(\ref{sch}). 
 
What then is the energy range for which a Klein region exists? 
The Klein region is defined by 
\begin{equation} 
\min{{\cal E}_+}<\varepsilon< \max{{\cal E}_-} 
\end{equation} 
Now, ${\cal E}_-$ is a monotonic decreasing function of $r$. It
attains its maximum on the horizon: $\max{{\cal E}_-}={\cal
E}_-(r=2M)=e^2/8M$. Its asymptotic value at infinity is $-\mu$.
${\cal E}_+$ has a positively diverging derivative on the horizon
where it attains the same value as ${\cal E}_-$, and an asymptotic
value at infinity, $\mu$, which it approaches with a positive slope.
This obviously implies that somewhere in the intermediate region
between the horizon and infinity, there is a point where ${\cal E}_+$
attains a minimum, {\it e.g.} a potential well. The location of this
minimum may be found by calculating the roots of the equation
$\partial_r {\cal E}_+=0$. There are two physical roots. One
determines the location of the potential minimum while the other
determines the location of the potential maximum. Now, the effect of
angular momentum may be ignored, since this will only strengthen the
potential barrier leading to a smaller particle creation rate. Doing
so, we may ask what are the allowed values of $\gamma\equiv e^2/8M
\mu$ for the existence of a Klein region. Obviously, for $\gamma=0$
($e=0$) which corresponds to the usual {\it neutral} Schwarzschild
spacetime, there is no level crossing, no Klein region, and thus no
particle production. This indicates that we should actually look at
the other extreme, where $\gamma$ is large.  However, for an electron
$e^2/(G\mu)\simeq 10^{15} {\textrm g}$. This means that for $\gamma$
to be large we must look at black-holes of mass not much bigger than
$10^{15} {\textrm g}$: the arena of mini black-holes. Keeping this in
mind, we find the minimum of the effective potential ${\cal E}_-$ for
large $\gamma$ which finally gives the energy range for the existence
of a Klein region:
\begin{equation} 
\mu\left [\sqrt{8\gamma-3\over 8\gamma-1} +{4\gamma\over 
(8\gamma-1)^2}\right ]<\varepsilon<{e^2\over 8M}. 
\end{equation} 
The function in the square brackets in the leftmost side of the 
inequality above tends rapidly to $1$ as $\gamma$ is increased (for 
$\gamma=1$ it is equal to $0.926$; for $\gamma=5$ it is equal to 
$0.987$). Therefore, we conclude that in practice, the energy range 
for the existence of a Klein region is 
\begin{equation} 
\mu<\varepsilon<{e^2\over 8M}. \label{range} 
\end{equation} 
This energy range is the same as the one in which the black-hole 
starts to superradiate (see the end of the previous section). 

It should be emphasized that unlike the case where the background
carries a definite sign of electric charge (like in the Kerr-Newmann
spacetime), here the background is {\it neutral}. Therefore, it would
seem that the black-hole should absorb statistically equal amounts of
particles and antiparticles. Hence, it should remain neutral. In
reality, a screening effect should take place. Since the point charge
which creates the electric field should repel particles (or
antiparticles) with the same sign of charge as its own, and attract
their counterparts, a cloud of charge would form around it.  This in
turn should screen the point charge and further lower the pair
creation rate. Furthermore, this charge segregation may alter the
probabilities for assimilation of particles (antiparticles) by the
black-hole, so that assimilation of particles with sign like the sign
of the test charge would be more probable. Thus, the black-hole may
become charged after all !
 
Unfortunately, calculating the corresponding {\it Debye length} of the 
problem is a formidable task and falls outside the scope of this work. 
Surely, the cloud of charge may be considered as a neutral gas 
consisting of charged particles which Coulomb-interact. Then, in order 
to solve the problem one must first solve Maxwell equations on the 
curved background with a source term representing the distribution of 
charge in the cloud. In the case of a completely ionized gas or 
plasma, this source term is given by a sum of Boltzmann's factors, one 
for each kind of particles, each multiplied by the charge carried by 
each kind of particles \cite{Landau}. However, this is valid only if 
the pairs may be considered to be thermally distributed, which is not 
generally true in the case at hand. 
 
\section{Where does the energy come from? - A speculation.} 
 
In the previous section it was shown that the system
particle-black-hole loses energy in the process of pair creation.  A
simple question may be raised -- what is the source of energy carried
by the pairs? The corresponding problem of the possible sources of
electromagnetic energy radiated away by an accelerated charge in flat
spacetime, troubled and still troubles scientists \cite{Harpaz} (it
was named as the ``Energy Balance Paradox''), and several answers were
suggested. Leibovitz and Peres \cite{Leibovitz} suggested that there
exists a charged plane, whose charge is equal and opposite in sign to
the accelerated charge, and that it recesses with the speed of light
in a direction opposite to the direction of the acceleration. The
interaction between this charged plane and the accelerated charge
supplies the energy carried by the radiation. Another suggestion, by
Fulton and Rohrlich \cite{Fulton}, is that the energy radiated is
supplied from the self-energy of the charge. In the problem considered
here there is a third energy source--the black-hole. Here, we show
that on thermodynamical grounds the option that the black-hole lose
mass during the process is not possible for massive black-holes, and
discuss the proposition that the self-energy of the test charge is the
energy source.
 
\subsection{Thermodynamical arguments} 
 
First assume that the black-hole loses energy given by $-\Delta M$.
This can be approximated by $-\Delta M\approx 2\varepsilon N$, where
$N$ is the number of pairs, and $\varepsilon$ is some mean energy
carried away by the pairs. In the nonrelativistic regime,
$\varepsilon\approx \mu$ where $2\mu$ is the pair rest-mass. In the
relativistic regime the rest-mass of the particle can be neglected, so
$\mu\ll\varepsilon$.  Now, since the black-hole is static and neutral,
its entropy is given simply by $S_{BH}=4\pi M^2$. Therefore, during
the process the black-hole changes its entropy by $\Delta S_{BH}=-8\pi
M \Delta M$.  Combining the two results, we find that
\begin{equation} 
\Delta S_{BH}=-16\pi M \varepsilon N. 
\end{equation} 
Now we make use of the generalized second law of thermodynamics (GSL). 
For the GSL to hold, the entropy carried out by the pairs must at 
least compensate for the entropy lost by the black-hole, namely 
\begin{equation} 
0\leq\Delta S_{\em world}=\Delta S_{\em BH} +\Delta S_{\em pairs}. 
\end{equation} 
Now, if the created particles are considered to be nonrelativistic, 
the entropy they carry is never far from the number of particles 
involved. Thus, 
\begin{equation} 
\Delta S_{pairs}\approx \eta N, 
\end{equation} 
where $\eta$ is a proportionality constant of order unity. The same is
known to be true in the other extreme when the created pairs are
assumed to be relativistic. For example, for black body radiation
$\eta=2\pi^4/(45 \zeta(3))\approx 3.6$, where $\zeta(z)$ is the
Riemann zeta function\cite{Morse}. Similarly, if the duration of the
pair production process is long, so that the pairs are allowed to
thermalize, they may be considered as particles obeying
Fermi-statistics with vanishing chemical potential\cite{Landau}. The
specific entropy is then $\eta=S/N=14\pi^4/(135 \zeta(3))\approx 8.4$.
Substituting $\Delta S_{BH}$ and $\Delta S_{\em pairs}$ into the GSL
we find
\begin{equation} 
 -16\pi\varepsilon M N+\eta N\geq 0. 
\end{equation} 
Hence 
\begin{equation} 
\mu M\leq\varepsilon M \leq {\eta\over 16\pi}=O(1). 
\end{equation} 
The conclusion must be that in the $1\ll \mu M$ regime, the black-hole
cannot be the dominant energy source; the black-hole is just too cold!
Accordingly, the energy must come from somewhere else.  A further
conclusion is that the black-hole may not lose entropy during the
process: the black-hole is involved in an adiabatic process making its
horizon area invariant \cite{disturbing,Mayo3}.
 
Actually, in the last derivation we have implicitly assumed that the
pairs are produced at a large distance from the horizon. It turns out
that the GSL even strongly forbids pair production by the black-hole
taking place at close proximity to the horizon. There the energy of
the pairs as measured locally is dominated by the electrostatic
self-energy, and in fact diverges.  To see this we note that if the
constituents of a particle-antiparticle pair are considered to be
quasistatic, then their conserved energy as measured at infinity,
$\varepsilon$, amounts to energy invested in rest-mass plus
electrostatic self-energy:
\begin{eqnarray} 
\varepsilon&=&(1-{2M/ r})\varepsilon_{\em local} \nonumber \\ 
&=&(1-{2M/ r})\left(\mu+ {q^2M\over 2r(r-2M)}\right), 
\end{eqnarray} 
where $r$ is the Schwarzschild coordinate of the particle (see
Eq.(\ref{energy_inf}) below). The first striking thing apparent from
the expression for $\varepsilon_{\em local}$ is that it diverges as
the particle approach the horizon, $r\rightarrow 2M$. Therefore, if
the pairs are located in the close proximity of the horizon then the
dominant part of their energy lies in electrostatic self-energy. In
the other extreme, when the particles are located far away from the
horizon, their electrostatic self-energy is small compared with the
rest-mass energy, until it vanishes at infinity. The point of
transition, $\tilde r$, from electrostatic self-energy dominated
region to rest-mass dominated region is set by the condition
\begin{equation} 
\mu={q^2 M\over 2\tilde r(\tilde r-2M)}.
\label{restmass_elec} 
\end{equation} 
Thus $\tilde r=2M f(e^2/8M\mu)$, where $f(x)=\sqrt{x+1/4}+1/2$. Note
that $f(0)=1$ - no electrostatic self-energy.  As was assumed in the
previous section, the pair production rate is exponentially
small. Therefore in order to produce a non-negligible number of pairs,
the exponent must be of order unity, {\it e.g.} we are limited by the
condition $1\lesssim e^2/M\mu$ (see Eq.~(\ref{range})). Taking this
into consideration we note that $f(1)\approx 1.62$. Hence, in our
approximation, the electrostatic self-energy is the dominant part of
the particles energy only in a narrow region around the horizon of
width, $\tilde r-2M\sim 0.62\times 2M$. Pair production in that region
is highly improbable on account of the GSL.
 
\subsection{Dynamical arguments} 
 
Realizing that the black-hole may not be the major energy source, we
turn now to look at the dynamics of the test charge as a second
candidate. We begin by considering the motion of a test particle of
mass $m$ and charge $e$.  Its motion, were it subject only to
\textit{gravitation and electromagnetic influences}, would be governed
by the Lagrangian
\begin{equation} 
 L = -m\int \sqrt{-g_{\alpha\beta} \, u ^\alpha \, u^\beta}\ d\tau + 
  e\int  A_\alpha \,u^\alpha d\tau, 
\label{lagrangian1} 
\end{equation} 
where $x^\alpha(\tau)$ denotes the particle's trajectory, $\tau$ the
proper time, and $u^\alpha=\dot x^\alpha=dx^\alpha/d\tau$, and $
A_\alpha$ means the background electromagnetic 4--potential evaluated
at the particle's spacetime position.  Recalling that
$g_{\alpha\beta}u ^\alpha u^\beta = -1$, it follows from the
Lagrangian that the canonical momenta are $p_\alpha=\delta {\cal
L}/\delta u^\alpha= m g_{\alpha\beta} \,u^\beta + e A_\alpha$.  The
stationarity of the envisaged background means there is a timelike
Killing vector $\xi^\alpha=\{1,0,0,0\}$.  The quantity
\begin{equation} 
\varepsilon\equiv -p_\alpha \xi^\alpha = -m g_{t\beta}\,u^\beta -e 
A_t , \label{energy} 
\end{equation} 
corresponds to the usual notion of energy as measured at infinity. Its
first term expands to $m +{1\over 2} m (d{\bf x}/dt)^2$ in the
Newtonian limit. The second term, $-e A_t$, is thus the electric
potential energy.

Varying the Lagrangian Eq.~(\ref {lagrangian1}) with respect to $ 
u^\beta$ gives the equation of motion of the particle 
\begin{equation} 
m{D u^\alpha\over d\tau}=e F^\alpha_\beta u^\beta, 
\label{motion} 
\end{equation} 
where $F_{\alpha\beta}=A_{\beta ;\alpha}-A_{\alpha ;\beta}$.  The
proper time derivative of $\varepsilon$, may be calculated as follows:
\begin{equation} 
\dot\varepsilon=-{d\over d\tau}(\xi^\alpha p_\alpha) 
=-\xi^\alpha \left(m{D u_\alpha\over d\tau}+e {D A_\alpha\over 
d\tau} \right)-e A_\alpha{D\xi^\alpha\over d\tau} . 
\end{equation} 
where we have used the fact that the proper time derivative of the
timelike Killing vector, $\xi^\alpha$, along the trajectory of the
particle is always perpendicular to the trajectory, ($u_\alpha
{D\xi^\alpha/ d\tau} =0$).

Now, If the test charge is assumed to be quasistatic (supported by
some mechanical apparatus), then its 4-velocity is given by
$u^\beta\approx ((-g_{tt})^{-1/2},0,0,0)$. Substituting for $A_\alpha$
from Eq.~(\ref{self}) into Eq.~(\ref{energy}) we find
\begin{equation} 
\varepsilon=m (1-2M/r) +e^2 M/(2r^2).
\label{energy_inf} 
\end{equation} 

Making use of the equation of motion (\ref{motion}), it is easy to
show that $\dot\varepsilon$ vanishes. Even if we go beyond the
quasistatic approximation and assume that the particle has a small but
non-vanishing radial velocity, $\varepsilon$ is still conserved. Does
it mean that the system cannot radiate?  We intend to show now how the
equation of motion should be modified to account for the irreversible
processes involved in pair production.
 
First, we define a 4-momentum rate of radiation by ${\cal R}
u^\alpha$.  In the case where the radiation is electromagnetic ${\cal
R}$ is given by the relativistic generalization of the Larmor formula
${\cal R}=(2/3) e^2 a ^\alpha a _\alpha$; $a ^\alpha$ is the
acceleration. In the problem considered here, a simple relation
between ${\cal R}$ and the dynamics of the test particle is unknown.
However, one may approximate the rate of energy loss due to pair
production using methods to be discussed below.

${\cal R} u^\alpha$, being a loss, should be subtracted from the right
hand side of the equation of motion (\ref{motion}).  One might expect
that in this way we have correctly accounted for the momentum and
energy loss due to pair production. Unfortunately, this equation is
inconsistent with ${\cal R}$ being positive definite; multiplication
of the modified equation of motion by $u_\alpha$, and using the
normalization condition, $u^\alpha u_\alpha=-1$ yields ${\cal R}=0$!
A term must be missing. To supply it we write
\begin{eqnarray} 
m{D u^\alpha\over d\tau}&=&e F^\alpha_\beta u ^\beta-\Gamma^\alpha 
, \nonumber \\ \Gamma^\alpha &\equiv& {\cal R} u^\alpha+S^\alpha.
\label{mmmotion} 
\end{eqnarray} 
$\Gamma^\alpha$ here may be understood as a 'frictional force',
and $S^\alpha$ is to be specified below.
 
There remains the question of what is the origin of that friction? One
possible answer is to view the test particle as a {\it Brownian
particle} interacting with a quantum field assumed to be in the vacuum
state \cite{gour}. At zero temperature, even though the thermal
fluctuations are absent, the quantum field still possess vacuum
fluctuations. Obviously, the particle cannot keep accruing energy from
the fluctuations present in the surrounding environment. Therefore,
there should exist a mechanism for the particle to dissipate its
energy so that it reaches equilibrium with the environment. Treating
the quantum field as a classical stochastic variable it is possible to
take the approach of Langevin who suggested, early this century, that
the force exerted on the particle by the surrounding medium can
effectively be written as a `rapidly fluctuating' part and an
`averaged out' part which represents a frictional force experienced by
the particle. The presence of the frictional force implies the
existence of processes whereby the energy associated with the particle
is dissipated to the degrees of freedom corresponding to the
surrounding medium.
 
Multiplying Eq.~(\ref{mmmotion}) by $u_\alpha$ now yields 
\begin{equation} 
\Gamma _\alpha u^\alpha=0 \quad\Longrightarrow \quad {\cal
R}=S^\alpha u_\alpha.
\label{cond1} 
\end{equation} 
We have obtained a constraint over $S^\alpha$. It is reasonable to
assume that $S^\alpha$ is a function of the velocity and its
derivatives. Given that the variation of the velocity is small, we may
expand
\begin{equation}
S^\alpha={\cal C}_0 u^\alpha+{\cal C}_1a^\alpha+{\cal C}_2\dot
a^\alpha+{\cal C}_3 \ddot a^\alpha \cdots,
\end{equation}
where $\{{\cal C}_i\}$ are proportionality constants with
corresponding dimensions of ${\em energy}\times {\em time}^{i-1}$. The
first term has to vanish since it is already included in ${\cal
R}u^\alpha$. The second term may be accounted for by mass
renormalization by the rule, $m\rightarrow m+{\cal C}_1$ (see the
equation of motion (\ref{mmmotion})). It turns out that the proper
time derivative of the acceleration, $ \dot a^\alpha$, is the lowest
derivative of the velocity allowed. Accordingly we set $S^\alpha=
{\cal C}_2\, \dot a^\alpha$.

In conjunction with the constraint (\ref{cond1}), we must go beyond
the static approximation. For if the particle is static then $u_0
u^0=-1$, and ${\cal R}=u^0 S_0$. Thus , a straightforward calculation
of $\dot\varepsilon$ using the modified equation of motion
(\ref{mmmotion}), gives
\begin{equation} 
\dot\varepsilon=-\xi^\alpha \Gamma_\alpha=- (u_0 
 u^0 S_0+S_0)=0 ,
\end{equation} 
regardless of the properties of $S^\alpha$. Accordingly, we set 
\begin{equation} 
u^\alpha=\left((-g_{tt})^{-1/2}\left(1+g_{rr}\delta u^2/2\right),\pm\delta
u,0,0\right),
\end{equation} 
where the velocity correction, $\delta u$, is assumed to be {\it time
independent} and small. Calculating $\dot a^\alpha$ to the lowest
order in $\delta u$ and its derivatives, and substituting into the
constraint (\ref{cond1}), we obtain a first order, non linear,
ordinary differential equation for $\delta u(r)$, whose solution is:
\begin{equation} 
\delta u(r)=\pm\left[{2\over M{\cal C}_2}(-g_{tt})^2\right]^{1/4}\left(\int^r_\infty
r^2{\cal R}dr\right)^{1/4}.
\end{equation} 
$\delta u$ scales with $(1/M)^{1/4}$. Hence, for massive black-holes
$\delta u$ is small. Moreover, $\delta u$ is proportional to the $1/4$-th
power of the energy dissipated along the trajectory of the particle,
hence it depends on the {\it history} of the particle.  ${\cal R}$
must drop at least as fast as $1/r^4$ for $\delta u$ to converge as
$r\rightarrow \infty$. In fact, it was already assumed in
Sec.\ref{klein} that ${\cal R}$ is proportional to the transmission
probability through the Klein region which is exponentially
small. This however determines only the energy scale over which pair
production may take place. It does not set the functional dependence
of ${\cal R}$ at the position of the test particle.

To resolve this we turn to Schwinger's approach \cite{Schwinger}, who
showed that the probability for pair creation per unit time per unit
volume by a constant electric field is
\begin{equation} 
\left({q E\over \pi}\right)^2 \sum_{k=1}^\infty {1\over k^2} e^{-{k 
\pi\mu^2\over q E}} \sqrt{-g}. 
\label{LI2} 
\end{equation} 
Multiplying this with some mean energy carried by the pairs gives an
approximation for ${\cal R}$. Although (\ref{LI2}) was originally
derived by calculating the imaginary part of the effective Lagrangian
for the Dirac field of rest-mass $\mu$ and charge $q$ in a prescribed
constant electrostatic field $E$ in {\it flat space time}, we adapt
this result to the corresponding problem in {\it curved spacetime} by
substituting local expressions for the energy and electric field. We
use the formula for the repulsive self-force as measured by an
instantaneously comoving, freely falling observer, at the position of
the test particle \cite{Smith}, to define the effective electric field
$E$:
\begin{equation} 
F_{self}=e E={e^2 M \over r^3}. 
\end{equation} 
This repulsive force is peculiar to charged test particles. Since the
black-hole is uncharged, this must be interpreted as arising from the
test particle's electrostatic self-interaction. It vanishes as $M$
vanishes, indicating that the effect is induced by the black-hole's
spacetime curvature.

Substituting for the effective electrostatic field in the formula for
the rate of pair production, Eq.~(\ref{LI2}), we find that ${\cal R}$,
vanishes exponentially fast as $r\rightarrow\infty$. Finally, putting
everything together we find that
\begin{equation}
{d\varepsilon\over d\tau}=-{\cal C}_2 u_0 u^r {D a_r\over d\tau}
=O({\cal R}^{5/4}),
\end{equation}
hence, the change in the energy of the test particle is negligiblly
small regardless of the sign of the velocity.
 
To summarize, as the particle is slowly lowered towards the black-hole
(or pulled back) by the mechanical apparatus, additional work must be
done against the frictional force induced by vacuum fluctuations. The
extra energy invested in moving the particle is then dissipated away
as pairs of massive charged particles.
 
\section{Summary and Assessment} 
 
It was demonstrated how Hawking's radiance form an isolated neutral
black-hole is modified on account of the electrostatic
self-interaction of charged particles. Once the temperature of the
black-hole is high enough so that the lightest massive charged
particles are emitted, the thermal radiation of charged particles
emitted by the black-hole is increased with respect to the thermal
radiation of neutral particles with the same mass and spin. This is a
direct consequence of the inclusion of the self-interaction into the
analysis.

An interesting consequence of this conclusion is that an {\it
external} electrostatic potential (as opposed to the {\it internal}
electrostatic self-potential) can also be used to modulate Hawking
radiance. Provided that the external potential is analytic in the
neighborhood of the horizon, the thermal radiation is modulated by
varying the magnitude of the potential. The modulation factor has the
form $\exp(e A^{\em ext}_t(2M)/T_H)$, where the external potential,
$A^{\em ext}_t$ is evaluated at the horizon and $T_H$ is the Hawking
temperature (see Eq.~(\ref{ssys})).

The possibility to modulate the thermal emission from the black-hole
has some very interesting consequences. First, assume that one applies
an external repulsive electrostatic field, opposing the gravitational
pull of the black hole. Then the probability to propagate from the
interior to the exterior of the horizon for a charged particle with
energy below $e A^{\em ext}_t(2M)$, would be greater than the
probability for the inverse process. If now the applied electrostatic
field is attractive (acting in the direction of the gravitational
pull), then it would serves as a high pass filter, suppressing the
emission of charged particles with energy below $e A^{\em
ext}_t(2M)$. This enable us to control the energy range over which the
black-hole superradiate charged particles! The similarity between this
effect and the phenomenon of thermionic emission is manifest.
 
Another issue that deserve further investigation is the calculation of
the rate of pair production. It was assumed in Sec.\ref{klein} that
the transmission coefficient, $T^2$, through the potential barrier
separating the positive from the negative energy states is
proportional to the probability for an incident particle to create a
pair of particles.  The Klein region for the problem (including the
effect of self-interaction) was found and shown to correspond to the
energy range for black-hole superradiance of charged
particles. Thought $T^2$ may be calculated numerically, it would be
profitable if one could find an analytic approximation for $T^2$, and
show that the result converges to the result obtained using
Schwinger's approach \cite{Gibbons}. The problem is that, Schwinger's
approach was originally formulated in {\it flat spacetime}, and the
formulation of this approach in {\it curved spacetime} may very much
prove to be an uphill task. So that a way to bypass these difficulties
is much needed.

Finally, the problem of energy source for the pair production process
was discussed.  It was shown, that on thermodynamical grounds, it is
not possible for a massive black-hole to lose mass during the
process. This is just to say that the black-hole is too cold. More
precisely, the entropy outflow from the system is too low for the
generalized second law to hold. We thus turned to explore the dynamics
of the test charge as a second candidate. It was speculated that
vacuum fluctuations of a quantum field interacting with the test
particle may be involved in the process of pair production. These
vacuum fluctuation, induce random motion that the particle undergoes,
and an averaged-out force that enters into the equation of motion as a
friction term. This friction term is a manifestation of the
dissipation mechanism by which energy is given off in the form of
massively charged particles. That being the case, a relation between
the friction term and the rate of energy dissipation, was found. As
could be anticipated, a static system can not radiate. Accordingly,
going beyond the static approximation, it was assumed that the test
particle has a small (but non-negligible) radial velocity. Then, the
functional dependence of the particle velocity on the rate of energy
dissipation was determined. It was shown that the velocity scales
inversely with the black-hole mass and proportional to the $1/4$-th
power of the energy dissipated along the trajectory of the particle,
hence it depends on the {\it history} of the particle.

The picture that seems to arise is that as the particle is lowered
towards the black hole, or pulled away, the mechanical apparatus
supporting the particle is doing work in changing the particle's
energy to the value appropriate to the new location. However, as the
particle moves, it interacts with the vacuum fluctuations in the
medium which have the effect of inducing frictional forces. If the
particle is assumed to move in a constant velocity, then these
frictional forces must be overcome by the investment of additional
work on part of the mechanical support. The extra energy is then
dissipated away in a process of pair production.Obviously for the
establishment of this picture, further study of the relationship
between vacuum fluctuations in {\it curved spacetime} and friction is
in order.
  
{\bf ACKNOWLEDGMENTS} The author thanks J. Bekenstein for his valuable 
suggestions and advice, and L. Sriramkumar and M. Schiffer for many 
discussions. This research is supported by a grant from the Israel 
Science Foundation, established by the Israel Academy of Sciences and 
Humanities.

\end{document}